\documentclass[11pt,titlepage,final]{amsart}
\usepackage{epsfig}
\usepackage{axodraw}

\newcommand{\half}{\tfrac12}
\newcommand{\quarter}{\tfrac14}
\newcommand{\sixth}{\tfrac16}

\newcommand{\p}{\mathcal {P}}
\newcommand{\diff}{{\rm d}}
\newcommand{\h}{\mathcal {H}}
\newcommand{\N}{\widehat{N}}
\newcommand{\V}{\mathcal {V}}
\newcommand{\R}{\mathcal {R}}
\newcommand{\av}[1]{\langle{#1}\rangle}
\newcommand{\q}{\mathfrak{q}}
\newcommand{\ket}[1]{|#1\rangle}
\newcommand{\bra}[1]{\langle#1|}
\newcommand{\tba}{{\sc tba}}
\newcommand{\dn}{{\rm dn}}
\newcommand{\sn}{{\rm sn}}
\newcommand{\cn}{{\rm cn}}
\newcommand{\IR}{{\sf IR}}
\newcommand{\UV}{{\sf UV}}

\begin{document}
  \begin{titlepage}
    \begin{flushright}
      {\small \sf FERMILAB-Pub-03/165-T} \\
      {\small\sf LPM/03-013}\\
      {\small\sf UPRF-2003-12}
    \end{flushright}
    \vskip 1.in
    \begin{center}
      \textbf{\Large An eigenvalue problem related to the\\
        non-linear $\sigma$--model:\\[.2em] analytical and numerical
        results}\\[2.em] \textbf{\large V.A.\ Fateev\footnote{\it
          Laboratoire de Physique Math\'{e}matique, Universit\'{e}
          Montpellier II, Pl. E.\ Bataillon, 34095 Montpellier,
          France, {\sf fateev@lpm.univ-montp2.fr}}$^,$\footnote{On
          leave of absence from Landau Institute for Theoretical
          Physics, ul.Kosygina 2, 117940 Moscow, Russia.}}
      \hspace{0.1cm} {\large and} \hspace{0.1cm} \textbf{\large 
        E.\ Onofri\footnote{\it Fermi National Accelerator Laboratory,
          P.O. Box 500, Batavia, Illinois, 60510,
          USA.}$^,$\footnote{Permanent address: Dipartimento di
          Fisica, Universit\`a di Parma, {\small\sf I.N.F.N.}, Gruppo
          Collegato di Parma, 43100 Parma, Italy, {\sf
            onofri@unipr.it}} }\\[2.em]
    \end{center}
    
  \bigskip
  \bigskip
  \begin{center}
    \textbf{Abstract}
  \end{center}
  {\small An eigenvalue problem relevant for non-linear sigma model
    with singular metric is considered. We prove the existence of a
    non-degenerate pure point spectrum for all finite values of the
    size $R$ of the system. In the infrared (\IR) regime (large $R$)
    the eigenvalues admit a power series expansion around \IR\ 
    critical point $R\rightarrow\infty$. We compute high order
    coefficients and prove that the series converges for all finite
    values of $R$.  In the ultraviolet (\UV) limit the spectrum
    condenses into a continuum spectrum with a set of residual bound
    states. The spectrum agrees nicely with the central charge
    computed by the Thermodynamic Bethe Ansatz method.}
\end{titlepage}

\section{Introduction}\label{sec:introduction}
The non-linear sigma models in two-dimensional (2D) space time are
widely used in field theory as continuous models of two-dimensional
spin systems (see e.g. Ref.s.~\cite{polyakov75, brezin76, friedan80,
  zinn-justin89}) as well as in relation to string theory (e.g.
Ref.s.~\cite{fradkin82, cfmp85, lovelace84, candelas85}). The general
2D sigma model (SM) is defined through the action
\begin{equation}
\mathcal{A}[G]=\frac{1}{2}\int G_{ij}(X)\,\partial _{\mu }X^{i}\,
\partial _{\mu}X^{j}\,{\rm d}^2 x  \label{eq:generalsm}
\end{equation}
where the coordinates $x^{\mu },\,\mu =1,2$ span a 2D flat space-time,
while the fields $X^{i},\,i=1,...,d$ are coordinates in a
$d$-dimensional Riemannian manifold called target space. The symmetric
matrix $G_{ij}$ is the corresponding metric tensor.

The standard approach to 2D sigma models is  perturbation theory. If the
curvature of $G_{ij}$ is small, one can use the following one-loop
renormalization group equation from Ref.\cite{friedan80}
\begin{equation}
\frac{\diff}{\diff t}G_{ij}=-\frac{1}{2\pi }R_{ij}  \label{eq:rg}
\end{equation}
where $t$ is the RG ``time'' (the logarithm of scale) and $R_{ij}$ is the
Ricci tensor of $G$.

The simplest examples of 2D sigma models are SM with two-dimensional
target space ($d=2$). In this case $R_{ij}=\mathcal{R}\,\delta _{ij}$
where $\mathcal{R}$ is the scalar curvature. Then we can always choose
(at least locally) conformal coordinates for which
\begin{equation}
G_{ij}=e^{\Phi }\,\delta _{ij}  \label{cc}
\end{equation}
with a single function $\Phi $.

An important role in the analysis of 2D sigma models is played by the
effective central charge $c(R)$. This dimensionless function contains
the information about the \UV\ and \IR\ properties of the theory and
it is related to the ground state energy $E_{0}(R)$ of the
corresponding quantum system, living on a finite space circle of
length $R$:

\begin{equation}
E_{0}(R)=-\frac{\pi c(R)}{6\,R}.  \label{eq:cch}
\end{equation}
For SM with two-dimensional target space $c_{UV}=c(0)=2$.

In integrable theories this quantity can be calculated exactly using
Thermodynamic Bethe Ansatz (\tba) equations Ref.\cite{yang69,
  zamolodchikov90}.  This problem is however much more complicated for
the excited levels $E_{i}(R)$, so it is useful to have some
independent approach for their calculation.  It was shown in
Ref.\cite{fateev93} that for the sigma models with $d=2$, in one loop
approximation (Eq.\ref{eq:rg}), this problem can be reduced to the
eigenvalue problem for the operator:
\begin{equation}
\widehat{h}=-\frac{1}{2}\nabla _{t}^{2}+\frac{1}{8}\mathcal{R}_{t}\,,\qquad 
\widehat{h}\Psi _{i}=\frac{\pi e_{i}(R)}{6}\Psi _{i}.  \label{egvp}
\end{equation}
Here $\nabla _{t}^{2}=e^{-\Phi }(\partial _{x}^{2}+\partial _{y}^{2})$ is
the Laplace operator and $\mathcal{R}_{t}$ is the scalar curvature in the SM
metric renormalized at the scale $R$:
\begin{equation}
t-t_{0}=\log R\Lambda _{0}  \label{ttt}
\end{equation}
where $\Lambda _{0}$ is the normalization parameter.
This operator is self-conjugate with respect to the scalar product in the SM
metric:
\begin{equation}
(\Psi _{1,}\Psi _{2})=\int \Psi _{1}^{\ast }\Psi _{2}\,e^{\Phi }
\,\diff x \,\diff y.
\label{eq:spr}
\end{equation}
The effective central charge $c(R)$ in one loop approximation can be
expressed through the lowest eigenvalue :
\begin{equation}
c(R)=2-e_{0}(R)  \label{ceff}
\end{equation}
and the excited levels $E_{i}(R) = E_{0}(R)+ \pi (
e_{i}(R)-e_0(R))/6R$. We note that if in the \IR\  limit SM flows to the
critical point described by conformal field theory (CFT) then the
numbers $\Delta _{i}=(e_{i}(\infty ) - e_{0}(\infty ))/24 $ coincide
with conformal dimensions of the fields in this CFT.

The eigenvalue problem (Eq.~(\ref{egvp})) with the natural scalar
product provided by the metric $G_{ij}(X)$ can be applied to the
analysis of 2D sigma models with target space of arbitrary dimension
$d$ . It follows from Zamolodchikov's $c$-theorem that the effective
central charge defined in Ref.\cite{zamolodchikov86} through the
correlation functions of the energy momentum tensor is non-increasing
as a function of the scale $R$. The effective central charge defined
by Eq.~(\ref{ceff}) (with $d-e_0(R)$ in the r.h.s.) also satisfies
this remarkable property, which follows from one of the results of
Ref.\cite{perelman02}, where it was shown that the lowest
eigenvalue of the operator $\widehat{h}$ is a non-decreasing function
of $R$.

A well known solution to Eq.~(\ref{eq:rg}) (see Ref.\cite{fateev93})
defines the axially symmetric metric of the ``sausage'' SM (an
integrable deformation of the $O(3)$ non--linear sigma model), which
is described by the action:
\begin{equation}
\mathcal{A}_{ssg}=\int \frac{(\partial _{\mu }X)^{2}+
(\partial _{\mu }Y)^{2}}{a(t)+b(t)\cosh2Y}\,{\rm d}^{2}x  \label{eq:saus}
\end{equation}
where $a(t)=\nu \coth 2u,~b(t)=\nu /\sinh 2u,$ and $u=\nu (t_{0}-t)/4\pi $.
It is easy to see from the explicit form of the metric (\ref{eq:saus}) that
operator $\widehat{h}/\nu $ depends only on the variable $u$ and does not
depend on parameter $\nu $. It means that
\begin{equation}
e_{i}^{ssg}(R)=\frac{\nu }{4\pi }\kappa _{i}^{ssg}(u).  \label{sc}
\end{equation}
where after the substitution $\Psi =\exp (imx)\Psi _{m}(y)$ ($m\in
\mathbb{Z} $) the scaling function $\kappa _{i}^{ssg}(u)$ is the
eigenvalue of the Sturm-Liouville problem:
\begin{equation}
\left[ -\partial _{y}^{2}+m^{2}+\frac{1+\cosh 2u\cosh 2y}{(\cosh 2u+\cosh
2y)^{2}}-\frac{\sixth\kappa _{i}^{ssg}(u)\sinh u}{\cosh 2u+\cosh 2y}\right]
\Psi _{m}^{(i)}=0  \label{eq:slp}
\end{equation}
where the eigenfunctions $\Psi _{m}^{(i)}$ have finite norm according
to Eq.(\ref{eq:spr}). For the ground state $\kappa _{0}^{ssg}(u)$ this
problem was studied in Ref.\cite{fateev93}.

In this paper we consider the eigenvalue problem for the sigma model which
correspond to another solution of RG Eq.~(\ref{eq:rg}). This solution can be
obtained by analytic continuation $Y\rightarrow Y+i\pi
/4,~u\rightarrow u+i\pi /4$ from the solution for the sausage model. The
corresponding action can be written as:
\begin{equation}\label{eq:SM}
\mathcal{A}=\int \frac{(\partial _{\mu }X)^{2}+(\partial _{\mu }Y)^{2}}
{\alpha (t)+\beta (t)\sinh2Y}\,{\rm d}^{2}x  
\end{equation}
where $\alpha (t)=\nu \tanh 2u,~\beta(t)=\nu /\cosh 2u,$ and $u=\nu
(t_{0}-t)/4\pi $.  This metric has a singularity at $Y=-u$. It means
that the coordinate $Y$ in target space should be considered only in
the region $Y>-u.$ The curvature $\mathcal{R}$ also has singularity at
this point. However for small values of parameter $\nu $ the curvature
is not small only in the narrow region ($\delta Y\sim \nu )$ in the
vicinity of singularity. A more careful analysis shows that the one
loop approximation is valid for the calculation of the observables in
SM (\ref{eq:SM}). The relative correction to one loop approximation as
well as in the sausage SM has the order $\nu \log (1/\nu )$.  The
eigenvalue equation for the scaling functions $\kappa _{i}(u)$ ($
e_{i}(R)=\nu \kappa _{i}(u)/4\pi $) has now the form:
\begin{equation}
\left[ -\partial _{y}^{2}+m^{2}-\frac{1-\sinh 2u\sinh 2y}{(\sinh 2u+
\sinh2y)^{2}}-\frac{\sixth\kappa _{i}(u)\cosh u}{\sinh 2u+\sinh 2y}\right] 
\Psi_{m}^{(i)}=0 \label{eq:bp}
\end{equation}
The solution $\Psi _{m}^{(i)}(y)$ should satisfy now the boundary condition
\begin{equation}
\Psi _{m}^{(i)}(y) \underset{y\rightarrow -u }\approx (y+u)^{1/2}
\label{bc}
\end{equation}
and it must be square integrable with respect to the natural norm
\begin{equation}
\Vert\Psi _{m}^{(i)}\Vert^2 = \int_{-u}^{\infty }
\frac{\mid\Psi_{m}^{(i)}(y)\mid^{2}}{\sinh 2u+\sinh 2y}\,\diff y\;.
  \label{norm}
\end{equation}

In the IR limit $u\rightarrow -\infty $ the metric (\ref{eq:SM}) has
an asymptotic which can be written in the form of Eq.~(\ref{cc}) with
$\exp\{-\Phi _{IR}\} = \half\nu(\exp\{2Z\} - 1)$, with $Z=Y+u$.  For
discrete values of the parameter $\nu =4\pi/N$ the SM with this metric
can be derived from the $SU(2)$ level $N$ WZW models by gauging $U(1)$
symmetry (see Ref.\cite{kiritis91, maldacena01} for details).  The
resulting $ SU(2)_{N}/U(1)$ coset model described by the SM with
metric $\exp (\Phi _{IR})\delta _{ij}$ coincides with $Z_{N}$
parafermionic CFT of Ref.\cite{fateev86}. For general values of $u$
the quantum field theory (QFT) corresponding to SM~(\ref{eq:SM}) can
be considered as the deformation of parafermionic CFT. It is natural
to expect that it will be a massless theory describing the RG flow
from rather non-trivial UV field theory (which is also well defined
for the same discrete values of parameter $\nu $) with $c_{UV}=2$ to the
parafermionic CFT with $c_{IR}=2-6/(N+2)$ in the IR limit.  The
scaling functions $\kappa _{i}(u) $ in this case describe the RG
dynamics of energy levels from \UV\ regime to \IR\ asymptotics, where
they define (with relative accuracy $O(1/N)$) the spectrum of
anomalous dimensions of parafermionic CFT (see Appendix C).

This massless QFT is integrable and can be described in terms of
factorized scattering theory for massless excitations. For $\nu
=4\pi/N$ with $N\geq 3$ the ground state energy Eq.(\ref{eq:cch}) of
the SM (\ref{eq:SM}) (as well as that of the sausage
SM~(\ref{eq:saus}), of Ref.\cite{fateev93}) admits an exact calculation
by \tba\ method. These equations will be described in
Sec.~\ref{sec:matching-smallsf-tba}.

Both eigenvalue problems Eq.(11,13) are believed to have a purely
discrete spectrum. The ground state eigenvalue $\kappa _{0}^{ssg}$ of
Eq.~(\ref{eq:slp}) was studied in Ref.\cite{fateev93}, where the
asymptotics of this function in the regimes $u\rightarrow 0$ and
$u\rightarrow \infty $ were found. This function was studied
numerically in Ref.\cite{brunelli95} and the result was in perfect
agreement with the scaling function calculated from \tba\ equations
(see Sec.5).  Actually it is rather easy today to attack the spectral
problem Eq.~(\ref{eq:slp}) by using sparse matrix techniques available
in mathematical libraries. Good accuracy can be achieved by
introducing a multi-grid discretization and using Richardson
extrapolation.  This approach, however, is not immediately applicable
to new equation (\ref{eq:bp}) due to its singular nature.  In the
following we shall bring the equation to a form which is suitable for
a detailed perturbative analysis (Sec.~\ref{sec:Heun}), to a second
form which allows an accurate asymptotic analysis in the \UV\ regime
(Lam\'e form, Sec.~\ref{sec:lame-formulation}), as well to a third
form, more suitable for a purely numerical approach
(Sec.~\ref{sec:numerical-solution}). Finally
(Sec.~\ref{sec:matching-smallsf-tba}) we shall exhibit the matching of
the ground state of Eq.~(\ref{eq:bp}) with the central charge of the
modified \tba\ system whose structure is given in Fig.~8. The
interested reader will find some further mathematical details in the
Appendices.
\section{The  connection with Heun's equation}\label{sec:Heun}
To bring Eq.~(\ref{eq:bp}) to a more manageable form,
we begin by re-absorbing the the integration measure into the wave
function. By defining 
\begin{eqnarray}
  \Psi(y)&=& \sqrt{\rho(y)}\,\phi(y)\,,\notag\\
  \rho(y) &=& \sinh\,2y + \sinh\,2u
  \label{eq:rho}
\end{eqnarray}
we find
\begin{equation}
  \label{eq:second}
  -\dfrac{\diff}{\diff y}\left(\rho(y)\dfrac{\diff\phi(y)}{\diff y}\right) +\left(m^2\rho(y)
   - \sinh2y \right)\,\phi(y) = \sixth\kappa\cosh2u\;\phi(y)
\end{equation}
Putting $x=e^{-2(y+u)}$ and $w=-e^{-4u}\,,$  Eq.~(\ref{eq:second}) is
transformed into the following
\begin{equation}   \label{eq:chi}
  \phi^{\prime\prime}(x) +
  \left(\dfrac{1}{x-1}+\dfrac{1}{x-w}\right)\,\phi'(x)+
  \dfrac{(1-m^2)x^2 -\sixth\kappa\,(1-w)x +
  m^2+w}{4x^2(x-1)(x-w)}\,\phi(x)=0
\end{equation}
of the Fuchsian type. A further substitution $\phi\rightarrow\surd
x\,f(x)$ reduces Eq.~(\ref{eq:chi}) to the form
\begin{equation}\label{eq:heun}
  f''(x)+\left(\dfrac{m+1}{x}+\dfrac{1}{x-1}+\dfrac{1}{x-w}\right)\,f'(x) + 
  \dfrac{(1+m)x- \q}{x(x-1)(x-w)}\,f(x)=0
\end{equation}
where the so--called {\sl accessory parameter\/} $\q$ is given by
\begin{equation}\label{eq:q}
  \q = \tfrac1{24}\left(\kappa\,(1-w) + 6(1+w)(1+2m)\right)\,.
\end{equation}
 This equation was analyzed by Heun in 1888 (see
Ref.~\cite{heun88, ince56}) who considered a general linear differential
equation of the second order with four Fuchsian singularities.  In
Heun's notation, the solution is formally given by
$F(w;\q;1+m,1,1+m,1;x)$, but this is of little use in practice. We
gain some insight from the fact that the limit $w\to -\infty$ is a
case of \emph{confluence} of singularities which takes us back to the
hypergeometric equation (see Ref.~\cite{ince56} for a general
treatment).
\subsection{The  eigenvalue problem in algebraic form} \label{sec:eigenv-probl-algebr}
It is well known that series solutions for Heun's equations can be
most conveniently constructed using the basis of Jacobi polynomials
$\,P_n^{(m,\,0)}(1-2x)\,$ (see e.g. \cite{Bateman55}, Vol.III).  We are
now going to show how to solve the eigenvalue problem by exploiting
this favorable basis: the problem will reduce to finding the spectrum
of an (infinite--dimensional) tridiagonal matrix for which efficient
algorithms are well--known to exist \cite{golub96}.  Let us consider
Eq.~(\ref{eq:heun}): after setting $\,y=1-2x\,$, we can easily expand the
solution in a series of Jacobi polynomials $\p_n^m\equiv P_n^{(m,\,
  0)}(y)$ by converting the differential equation in the form
\begin{equation}  \label{eq:alg}
  \h\,f\equiv\left\{(1-2w-y)\,\N(\N+m+1) + (1-y^2)\,\frac{\diff}{\diff y} -
  (1+m)\,y \right\}f = (2\q-m-1)\,f
\end{equation}
where $\N \p_n^m = n\,\p_n^m$. Now we can use the basic
properties of Jacobi's polynomials (see e.g. ~\cite{Bateman55},
Vol.II, \cite{luke69})
to reduce the operator $\mathcal{H}$ to the form 
$\mathcal{H}\,f=(1-2w)\,\mathcal{H}_0\,f + \V\,f$ whose action on the basis vectors is particularly simple:
\begin{equation}\label{eq:matrix}
  \begin{split}
    \mathcal{H}_0\, \p_n^m &= n\,(n+m+1)\,\p_n^m\\
    \V\,\p_n^m &= 
      m\dfrac{(2+m)n^2+(m+1)(m+2)n+m(m+1)}{(m+2n)(m+2n+2)}\,\p_n^m\\
    &-2\dfrac{n^2(m+n)^2}{(m+2n)(m+2n+1)}\,\p_{n-1}^m
    -2\dfrac{(n+1)^2(m+n+1)^2}{(m+2n+1)(m+2n+2)}\,\p_{n+1}^m 
  \end{split}
\end{equation}
We can now conveniently study the spectrum of $\q$ using this
tridiagonal matrix representation, by applying, for instance, the
technique of {\sl Sturm sequences and bisection\/} \cite{golub96}.  The
matrix representation also lends itself to a very simple perturbation
series expansion, as we discuss in the next section. We shall have to
refer to the matrix representation of $\h$ in the orthonormal basis
$\phi_n^m = (-)^n\sqrt{\frac{1}{2n+m+1}}\,\p_n^m$ as $V_{nn'}$:
\begin{eqnarray}
  V_{n+1,\,n} &=& \frac{2(n+1)^2(n+m+1)^2}{(2n+m+2)\sqrt{(2n+m+2)^2-1}}\\
  V_{n-1,\,n} &=& \frac{2n^2(n+m)^2}{(2n+m)\sqrt{(2n+m)^2-1}}\nonumber
\end{eqnarray}
diagonal terms being unchanged, but the former $\V$, being rational in
its indices, is more convenient for the calculation of perturbative
coefficients; we show in Appendix A, that we can use $\V$ without
modifying the standard algorithm.

\subsection{Perturbation theory.}\label{sec:pert-expans}
The tridiagonal matrix representation of $ \mathcal{H}$, when rewritten as $
\mathcal{H} = \varepsilon^{-1}(\mathcal{H}_0 + \varepsilon \mathcal{V})$, can be
used to calculate a perturbative expansion in the parameter
\begin{equation}
\varepsilon = (1-2\,w)^{-1}=(1+2e^{-4u})^{-1}\,.
\end{equation}
The convergence of the expansion is governed by the Kato-Rellich
theorem\footnote{See for instance \cite{kato95, reed78}}: let there
exist constants $a,b$ such that $\Vert \mathcal{V}\phi\Vert\le
a\Vert\phi\Vert+b\Vert\mathcal{H}_0\phi\Vert$ ($\Vert .\Vert$ denotes
$L_1$-norm).  Then the perturbative expansion defines a regular
analytic function for $|b\varepsilon|<1$.  For a tridiagonal matrix
it's not so difficult to find norm estimates; in our case it's simple
algebra to check that {\sl the column sums of the matrix elements of
  $\mathcal{V}$ coincide with the diagonal matrix elements of
  $\mathcal{H}_0$, up to an additive constant}, therefore we have
\begin{equation}
  \mathcal{V} = \mathcal{S}\,\left(\mathcal{H}_0 + (m+1)\mathbb{I}\right)
\end{equation}
with $\mathcal{S}$ a stochastic matrix\footnote{{\emph i.e.}
  $\sum_i\mathcal{S}_{i,j}\equiv 1$} and $\mathbb{I}$ the identity
matrix. Hence it follows
\begin{equation}
  \Vert\mathcal{V}\phi\Vert = \Vert \mathcal{S} (\mathcal{H}_0+(m+1)\mathbb I)\phi\Vert \le
  \Vert\h_0\phi\Vert + (m+1)\Vert\phi\Vert
\end{equation}
which implies that the perturbative series will converge for
$\varepsilon<1$, that is {\sl for all $u$}. Actually we can say more:
let us denote by $R(H,\mu) = (H + \mu)^{-1}$ the resolvent operator;
by applying Schur's Test to the matrix $R(\h_0,\mu)\, V$ (the
symmetric version of $\V$) one concludes that
\begin{equation}
  \Vert R(\h_0,\mu)\,V \Vert \le 1
\end{equation}
if $\mu\ge m+13/8$. Now the resolvent of $\h$ satisfies
Lippman--Schwinger equation
\begin{equation}
  R(\h,\mu) = R(\h_0,\mu) - \varepsilon R(\h_0,\mu)\,V\,R(\h,\mu)
\end{equation}
which for $|\varepsilon|<1$ can be inverted to give
\begin{equation}\label{eq:LS}
  R(\h,\mu) = \left(1+\varepsilon R(\h_0,\mu)\,V\right)^{-1}\,R(\h_0,\mu)\;.
\end{equation}
Since $R(\h_0,\mu)$ is a compact operator, Eq.~(\ref{eq:LS}) is
telling us that \emph{the resolvent $R(\h,\mu)$ is itself a compact
operator\/}, being the product of a compact operator with a bounded
one. This implies that the spectrum of $\h$ is purely discrete.

The expansion can now be computed rather easily by the standard
recursive algorithm (see Appendix A).  Details on the series expansion
can be found in Appendix B, where we prefer to adopt a different
parameter which naturally appears in the Lam\'e formulation of next
section, namely
\begin{equation}\label{eq:lambda}
\lambda = (1+e^{-4u})^{-1} = \frac{2\varepsilon}{1+\varepsilon}\,,\; \varepsilon=\frac{\lambda}{2-\lambda}\,.
\end{equation}
The expansion in powers of $\lambda$ turns out to be simpler and with
better convergence properties; indeed the substitution $\varepsilon\to\lambda$
is just a special case of Euler's $(E,q)$-method
\cite{Hardy}. We present just a sample of the infinite number of
different series expansions, since we believe that nobody would like
to copy them from paper but would rather prefer getting the code which
generated the expansion\footnote{Matlab and Mathematica codes are
  available at the web site {\sf
    www.fis.unipr.it/{$\scriptstyle{\sim}$}onofri}.}.  The first few
terms for the ground state value of $\kappa$ at fixed $m$ are the
following
\begin{eqnarray}\label{eq:kappa}
\sixth \kappa_{m,\, 0} &=& 1+2m - \frac{2m}{m+2}\,\lambda - \frac{4
(m+1)^{3}}{(m+2)^{3}(m+3)}\,\lambda^{2} -  \\
&&-\frac{8(m+1)^{3}(2m^{2}+5m+4)}{(m+2)^{5}(m+3)(m+4)}\,\lambda ^{3} 
 + O(\lambda^4)\nonumber
\end{eqnarray}
while for the excited states, after putting $j=\half m+n$, we have
(see also Appendix C)
\begin{eqnarray}
\sixth\kappa_{m,\,n} &=&(2j+1)^{2}-m^{2}-\frac{\left[
4j(j+1)-m^{2}\right] ^{2}}{8j(j+1)}\, \lambda  - \\
&&-\frac{1}{2^{9}(2j+1)}\left[ \frac{(4(j+1)^{2}-m^{2})^{4}}{(j+1)^{3}(2j+3)}-\frac{(4j^{2}-m^{2})^{4}}{j^{3}(2j-1)}\right]\,\lambda ^{2} + O(\lambda^3)\;.\notag
\label{eq:jj}
\end{eqnarray}

Using floating point arithmetic we may quickly explore very high
orders, with due attention to truncation errors which accumulate along
the iteration.  The asymptotic behavior of the coefficients shows very
clearly a limit $c_{n+1}/c_n \to 1$, confirming that the series
converges in the unit circle, which means in the domain
$|e^{4u}/(1+e^{4u})| < 1$. In the complex $u$ plane this is a domain
which includes the whole real axis. However as we venture along the
positive real axis, the convergence is critically slowed down: to go
deep in the \UV\ region we may be obliged to sum a really huge number
of terms, or try some resummation, {\sl e.g.} via Pad\'e approximants.
Since high order coefficients are easily computed, however, we may try
to extract the asymptotic behavior of $\kappa(u)$ for large positive
$u$ by analyzing the asymptotic behavior of the coefficients.  For
example we can verify that the \UV\  asymptotics of the ground state
eigenvalue $\kappa _{0}(u)$ coincides with that of function $\kappa
_{0}^{ssg}(u)$ (see Sec.~\ref{sec:lame-formulation}) and has a form:
\begin{equation}
\kappa _{0}(u)=\frac{3\pi ^{2}}{2(u+\log 4)^{2}}+O(1/u^{5}).  \label{mrd}
\end{equation}
When expressed in terms of $\varepsilon$ this formula can be expanded
in a power series and the coefficients compared to those coming from
perturbation theory.  Sub-dominant terms tend to mask the simple
$n^{-1}(\log n)^{-3}$ behavior one should expect; we find that it is
more accurate to compare the Taylor coefficients of the function
$1/(4\q-1)$, which is actually diverging at $\varepsilon\to 1$, with
those of its leading term $O(\log({1-\varepsilon})^2)$.
However, to make the comparison even more transparent, we may look for
a special function whose behavior at $\varepsilon\to 1$ is the
simplest possible. Let us observe that $2\q-1$ turns out to be an {\sl
  odd} function of $\varepsilon$; the new function
$\sqrt{\varepsilon/(4\q-2+\varepsilon)}$ has the simple
$O(\log(1-\varepsilon))$ leading singular behavior at $\varepsilon\to
1$, and it is an {\sl even} function of $\varepsilon$. We argue that
its leading behavior should then be
\begin{figure}[ht] 
  \begin{center}
    \mbox{\epsfig{file=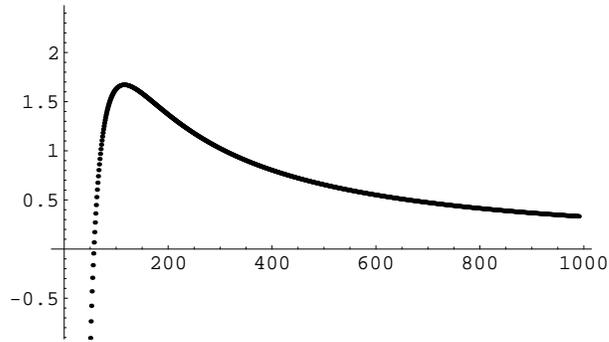,width=8.cm}}\label{fig:XI}
    \caption{\sl  Relative deviation of $\Upsilon$'s expansion coefficients from those of its leading asymptotics ($\times 10^5$) (Eq.\ref{eq:XI}).}
  \end{center}
\end{figure}
\begin{equation}\label{eq:XI}
  \Upsilon(\varepsilon)\approx \dfrac{1}{2\pi \varepsilon}\log\left(\dfrac{1+\varepsilon}{1-\varepsilon}\right)\;.
\end{equation}
In fact we find that the expansion of $\Upsilon$ matches the
perturbative series with a high accuracy (see Fig.~1 where the
deviation is magnified $10^5$ times).  The asymptotic behavior will be
recovered in a very precise way numerically in
Sec.~\ref{sec:numerical-solution}, hence its extraction from the
perturbative series appears of purely academic interest. Anyhow,
assuming Eq.~(\ref{eq:XI}) it follows from Eq.~(\ref{eq:q})
\begin{equation}
\kappa \equiv \dfrac{24\q-6(1+w)}{1-w}
\approx 6(4\q-1) = 6/\Upsilon^2
\approx \frac{3\pi^2}{2u^2}
\end{equation}
since
$\quarter\log((1+\varepsilon)/(1-\varepsilon))=\quarter\log(1+e^{4u})\approx
u$ as $\varepsilon\to 1$.
\subsection{Sausage model equation}\label{sec:sausage}
The sausage model was our starting point. Now we go back to it and
show how it fits into the correspondence with Heun's equation.  The
scaling function is defined in Eq.~(\ref{eq:slp}). We shall now look
for its algebraic equivalent as we did in Sec.~\ref{sec:Heun}.  Since
the two equations are related by analytic continuation, it will not
come out as a surprise that the differential equation is {\sl the
  same}, up to a map $w\to \bar w,\, \q\to \bar \q$. The range of
values for the problem of SM(\ref{eq:SM}) is $\quarter < \q < \half, w
< 0$, while the sausage is characterized by $\half < \bar \q < 1,\,
w>0$.  The point is that Eq.~(\ref{eq:slp}) can be brought to Heun's
form by the transformation $\xi = \exp\{2(y+u)\}.$ The singularities
are now located at $\{0, -1, -w,\, \infty\}$, with $w = \exp\{4u\}$
and $24 \q = 6(1+w) + \kappa(u)(w-1)$. The equation is actually {\sl
  the same} as the one we find for the SM(\ref{eq:SM}), but the domain
involved is the positive real line instead of the unit interval and
the singularities are differently situated. By applying well-known
transformation properties of Heun's equation (Ref.~\cite{heun88,
  kahmke59}) we can reposition the domain on the unit interval
$(0,1)$, the singular points being now $ 0, 1, \bar w = w/(w-1),
\infty$ and the new accessory parameter is given by
\begin{equation} \label{eq:acc}
  \bar \q = \frac{w-\q}{w-1}  \,.
\end{equation}
Notice that the map $\{\q,w\} \to \{\bar \q,\bar w\}$ is involutory with
$w=\infty$ as the only fixed point, the interval $\half< \bar \q< 1$
being mapped onto $1 < \q < \infty$. 

Hence we can use the same algorithm of the previous section in a
different domain. The ``magic'' here is provided by the analyticity
properties of the models involved. The complex shift transforming the
``sausage'' model into the SM~(\ref{eq:SM}) does not modify very much the
eigenvalue equation, which turns out to be the same equation in a
different domain.
\section{The Lam\'e formulation}\label{sec:lame-formulation}
The parameter $\lambda $ defined by Eq.(\ref{eq:lambda}) is naturally
related with a reformulation of Eq.~(\ref{eq:bp}) close to Lam\'e
elliptic equation.  If we define the modulus of Jacobi elliptic
functions
\begin{equation}
k^{2}=\lambda =1/(1+\exp (-4u))  \label{ellm}
\end{equation}
then the substitution
\begin{equation}
e^{y-u}=\frac{\dn(z|k^{2})}{k\, \sn(z|k^{2})}\;,\quad \psi _{m}=\sqrt{\frac{\sn(z|k^{2}) \dn(z|k^{2})}{\cn(z|k^{2})}}\Psi _{m}  \label{sbs}
\end{equation}
maps the point $y=\infty $ to $z=0$, the point $y=-u$ \ to $z=K$,
where $K(k^{2})$ is the real period of Jacobi elliptic functions, and
it turns Eq.~(\ref{eq:bp}) to the form:
\begin{equation}
\left( -\frac{\diff^{2}}{\diff z^{2}}-\frac{\dn^{2}(2z|k^{2})}{\sn^{2}(2z|k^{2})}+\frac{m^{2}\,\cn^{2}(z|k^{2})}{\sn^{2}(z|k^{2})\,\dn^{2}(z|k^{2})}\right) 
\psi _{m,\,n}=\sixth\kappa _{m,\,n}\,\psi _{m,\,n}  \label{du}
\end{equation}
with the boundary conditions $\psi _{m}\sim z^{m+1/2}$ at
$z\rightarrow 0;$ $\psi _{m}\sim (K-z)^{1/2}$ at $z\rightarrow K.$
This equation can be studied analytically in two limits
$k^{2}\rightarrow 0$ ($u\rightarrow -\infty $) and $k^{2}\rightarrow
1$ ($u\rightarrow -\infty $). In the first case we can develop the
standard perturbation theory near the exact solutions $\psi
_{m,\,n}(z)=\sqrt{\sin 2z}\,\cos ^{m}(z)\,P_n^{(0,\,m)}(\cos 2z)$
where $P_n^{(\alpha,\,\beta)}(x)$ are Jacobi polynomials. This
perturbation theory gives the same \IR\ expansion for the eigenvalues
which was considered in previous section.

In the opposite limit $u\rightarrow \infty $, $k\rightarrow 1$ and the
real period $K\sim -\frac{1}{2}\log ((1-k^{2})/16)\sim 2u+2\log
2\rightarrow \infty$. In this case the potential term in
Eq.(\ref{du}) is equal to $m^{2} $ almost everywhere and near the
points $z=0$; $z=K$ it can be approximated with exponential in $u$
accuracy by the potentials:
\begin{equation}
V(z)=-\frac{1}{\sinh ^{2}2z}+m^{2}\coth ^{2}z\qquad 0<z\ll K  \label{Vz}
\end{equation}
and
\begin{equation}
V_{1}(z_{1})=-\frac{1}{\sinh ^{2}2z_{1}}+m^{2}\tanh ^{2}z_{1}\qquad
0<z_{1}\equiv K-z\ll K  \label{V1}
\end{equation}
Both these potentials appeared in Ref.\cite{dijkgraaf92}, where the
spectrum of CFT describing Witten's two-dimensional Euclidean black
hole \cite{witten91} was studied.  There, it was noted that potential
$V_{1}(z_{1})$ is attractive and has the bound states solutions:
\begin{equation}
\psi _{m,\,n}\text{ }=\sqrt{\tanh z_{1}}\,(\cosh z_{1})^{2n-m+1}
\,F\left(-n,-n+m,\,m-2n;\,{1-\tanh^{2}z_{1}}\right)  \label{BS}
\end{equation}
where $F(a,b,c,z)$ is Gauss' hypergeometric function. These solutions are
normalizable for integer $n<(m-1)/2$ and give the levels 
\begin{equation}
  \label{eq:bs}
\sixth\kappa =\{ m^2 - (2n + 1 - m)^2 \;|\; n = 0,1,...,[\half m]-1\} \,.
\end{equation}
The corresponding eigenvalues of Eq.(\ref{du}) approach these levels
exponentially in $u$.

The potential $V(z)$ is repulsive and does not have normalizable
solutions. For $2n\geq m-1$ we parameterize $\kappa _{m,\,n}/6=$ $
m^{2}+p_{n}^{2}.$ Then solutions regular at $z=0$ and at $z_{1}=K-z=0$
are found to be
\begin{equation}
\psi _{m,\,n}(z)=(\tanh z)^{m+\frac{1}{2}}(\cosh z)^{ip}\;
F\left(\half(1+m-ip), \half(1+m-ip), m+1; \tanh^{2}z\right)  
\label{S1}
\end{equation}
\begin{equation}
\psi _{m,\,n}(z_{1})=(\tanh z_{1})^{\frac{1}{2}}(\cosh z_{1})^{ip}\;
F\left(\half(1+m-ip), \half(1-m-ip), 1; \tanh^{2}z_{1}\right)  
\label{S2}
\end{equation}
Matching these solutions with the plane wave solution in the region
$0\ll z\ll K$ we obtain the quantization condition:
$p_{n}=\quarter\pi(2n-m+2)/(u+r_{m})$ where $r_{m}=\psi (1)-\psi
(\half m+\half)$, and $\psi(x)$ is the logarithmic derivative of the
$\Gamma$ function. This condition leads to the asymptotics
\begin{equation}\label{eq:asympt}
\sixth\kappa_{m,\,n}(u) = m^{2}+\pi ^{2}\frac{(2n-m+2)^{2}}
{16(u+r_{m})^{2}}+O(1/u^{5});\quad n\geq \half(m-1) 
\end{equation}
(see Fig.~\ref{fig:UV}).  We note that for $m\neq 0$ this \UV\ 
behavior is different from that for the sausage model eigenvalues,
which is given by
\begin{equation}
\sixth\kappa_{m,\,n}^{ssg}(u) = m^{2}+
\pi^{2}\frac{(n+1)^{2}}{4(u+r_{m})^{2}}+O(1/u^{5});\quad (n\geq 0)
\label{uvss}
\end{equation}
The sausage model eigenvalue problem of Eq.~(\ref{eq:slp}) can also be
written in the elliptic form of Ref. \cite{fateev93} with
$k_{s}^{2}=1-\exp (-4u)$:
\begin{equation}
\left( -\frac{\diff^{2}}{\diff z^{2}} - \frac{\cn^{2}(2z|k_{s}^{2})}{\sn^{2}(2z|k_{s}^{2})} + 
\frac{m^{2}\,\dn^{2}(z|k_{s}^{2})}{\sn^{2}(z|k_{s}^{2})\,\cn^{2}(z|k_{s}^{2})}%
\right) \psi _{m,\,n}=\frac{\kappa _{m,\,n}^{ssg}k_{s}^{2}}{6}\psi _{m,\,n}
\label{ES1}
\end{equation}
with boundary conditions $\psi _{m}\sim z^{m+1/2}$ at $z\rightarrow 0;$ $%
\psi _{m}\sim (K-z)^{m+1/2}$ at $z\rightarrow K.$ In the \UV\  limit
$u\rightarrow \infty $ the potential term in this equation at the both
ends tends to $V$ given by Eq.(\ref{Vz}). The normalizable solutions
do not appear and both asymptotics can be described by the
eigenfunction (\ref{S1}). The quantization condition for parameter
$p_{n}$ in this case leads to the asymptotics Eq.(\ref{uvss}).

\section{Numerical analysis}\label{sec:numerical-solution}

The matrix representation introduced in a previous section, while
useful from the analytic viewpoint, is not the best choice if we want
to compute the spectrum beyond perturbation theory. Actually an
$n$--dimensional truncation of the matrix given in
Eq.~(\ref{eq:matrix}) is going to be essentially equivalent to $n$--th
order perturbation theory. We are now introducing another
transformation of Eq.~(\ref{eq:bp}) which will allow us to efficiently
explore the whole range $-\infty < u < \infty$.

We start from the fact that in the limit $u\to-\infty$ (and $m=0$)
there is a simple solution with
\begin{equation}
  \Psi(y) =  \sqrt{1-e^{-2(y+u)}}\equiv\sigma(y)\,,
\end{equation}
with $\kappa(u)\to 6$. Hence it seems promising to look for a
solution of the form
\begin{equation}
  \label{eq:ansatz}
  \Psi(y) = \sigma(y)\,\psi(y)
\end{equation}
To find the new differential equation it is convenient to write down the
functional $\av{\h}$ whose critical points are the eigenvalues: by denoting
$ \;\chi(y)\equiv 1-\sinh2u\,\sinh2y\; $
and recalling $\rho(y)$ from Eq.~(\ref{eq:rho}), we have
\begin{equation}  \label{eq:functional}
  \av{\h}  = \frac{\int_{-u}^\infty \left(\Psi(m^2-\partial^2)\Psi -
  \chi\rho^{-2}\right) \Psi^2)\,\diff y} {\int_{-u}^\infty
  \rho^{-1}\Psi^2\,\diff y}
\end{equation}
Now, by inserting Eq.~(\ref{eq:ansatz}), after an integration by part, 
we find 
\begin{equation}
  \label{eq:simplified}
  \av{\h} = \frac{\int_{-u}^\infty \left[\sigma^2\psi'^2 + \left(\sigma'^2
      -\chi\sigma^2/\rho^2 - (\sigma\sigma')' + m^2\,\sigma^2\right)
      \psi^2\right]\,\diff y} {\int_{-u}^\infty
      \sigma^2\rho^{-1}\psi^2\,\diff y}
\end{equation}
The variational equation resulting from this functional
\begin{equation}
  \label{eq:proto}
  -\frac{ \diff }{\diff y}\left(\sigma^2\,\frac{\diff \psi}{\diff y}\right) + V\psi =
   \sixth\kappa\,\cosh2u\;\sigma^2\,\rho^{-1}\,\psi
\end{equation}
suggests to introduce a new coordinate $\xi$ such that
$\diff y=\sigma^2\,\diff \xi$; we find
\begin{equation}
  y = -u + \half\log(1+e^{2 \xi}) \label{eq:newdefs}  
\end{equation}
and the equation simplifies to 
\begin{equation}\label{eq:VW}
  -\frac{\diff^2\psi}{\diff\xi^2} + \sigma^2\,V(\xi,u)\,\psi =
  \sixth\kappa(u)\,W(\xi,u)\,\psi(\xi)
\end{equation}
with 
\begin{eqnarray}
  \label{eq:potentials}
  \sigma^2V &=&
  \frac{e^{2\xi}((1+e^{4u})^2+(1+e^{2\xi})^2-1+e^{4u}(e^{4\xi}+3e^{2\xi}))}{(1+e^{2\xi})^2(1+e^{4u}+e^{2\xi})^2}
  + \quarter m^2(1+\tanh\xi)^2\\
  W &=&   \frac{e^{2\xi}(1+e^{4u})}{(1+e^{2\xi})(1+e^{4u}+e^{2\xi})}
\end{eqnarray}
The crucial property of this new formulation is that it is now {\sl
  regular on the whole real axis\/}.  With appropriate boundary
conditions (Neumann's b.c.)\footnote{It is somewhat tedious to trace
  the b.c. back from the original equation; suffices to say that,
  intuitively speaking, since the constant solution is exact in the
  limit $u\to-\infty$, Neumann b.c. are the natural ones.} the
eigenvalue equation can now be easily solved by standard
sparse--matrix generalized eigenvalue routines \footnote{We
  successfully used the routine {\sf eigs} in Matlab.}
\begin{figure}[ht] 
  \begin{center}
    \mbox{\epsfig{file=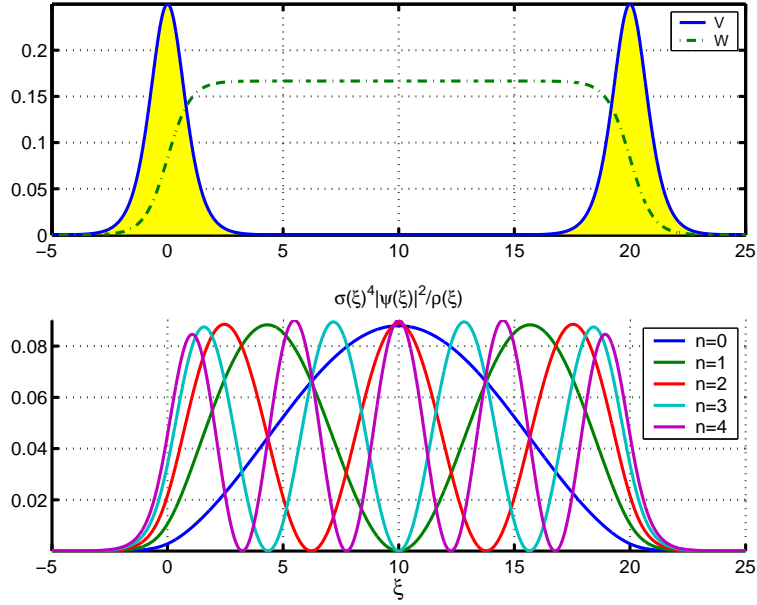,width=10.cm}}
    \caption{\sl The generalized potentials $V, W$ (upper) and the first
      eigenfunctions $\psi_n^2$ for $u=10, m=0$.}
    \label{fig:deepUV}
  \end{center}
\end{figure}
\begin{figure}[ht] 
  \begin{center}
    \mbox{\epsfig{file=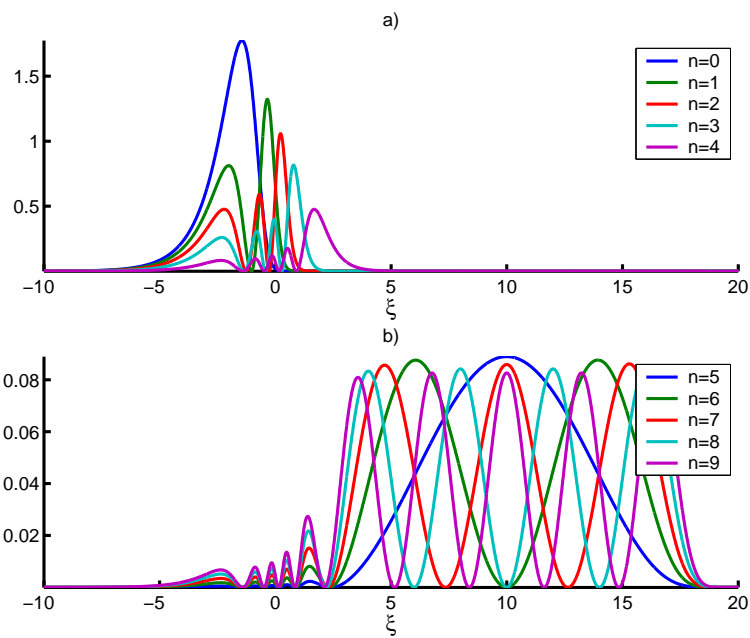,width=10.cm}}
    \caption{\sl The densities at $u=10, m=10$: a) ``bound
    states'', b) the ``continuum''.}
    \label{fig:mone}
  \end{center}
\end{figure}

\begin{figure}[ht] 
  \begin{center}
    \mbox{\epsfig{file=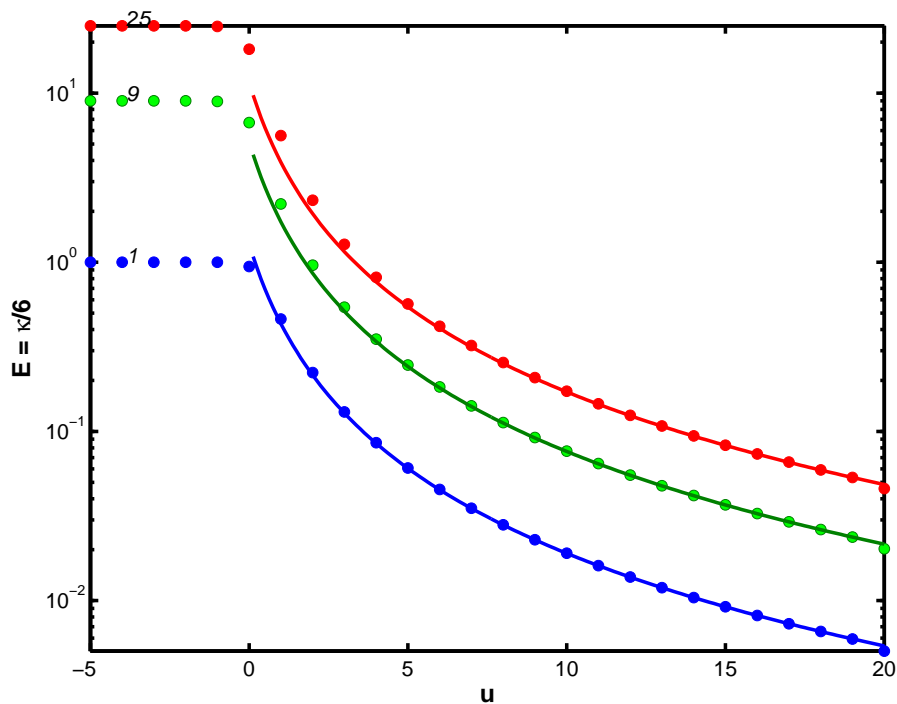,width=12.cm}}
    \caption{\sl The asymptotic behavior in the UV for the low lying
      states at $m=0$: lines $\to$ Eq.~(\ref{eq:asympt}), circles from
      numerical solution Eq.~(\ref{eq:VW}).}\label{fig:UV}
  \end{center}
\end{figure}

See Fig.s~(\ref{fig:deepUV},\ref{fig:mone}) for some
typical waveforms. The density includes the measure appropriate for
the new variable $\xi$, namely $\sigma(\xi)^4/\rho(\xi)$.
In the limit $u\to-\infty$ we easily recover the discrete spectrum 
\begin{equation}
  \label{eq:specIR}
\sixth\kappa = \{(2j+1)^2-m^2\,|\;j=n+\half m, n=0,1,2,\ldots\}  
\end{equation}
as we already know from perturbation theory (see
Eq.~(\ref{eq:jj})).  In the other limit, $u \gg 0$, the spectrum can
be described as a monotonously decreasing flow toward $\kappa/6=m^2$,
{\sl except for a finite number of eigenvalues} which have a value
less than $m^2$: these peculiar ``bound states'' are given by Eq.
(\ref{eq:bs}) and they are easily reproduced numerically (see
Fig.~\ref{fig:UV}). We easily check that they nicely agree with the
asymptotic formulas already given in Eq.~(\ref{uvss}).
\begin{figure}[ht] 
  \begin{center}
    \mbox{\epsfig{file=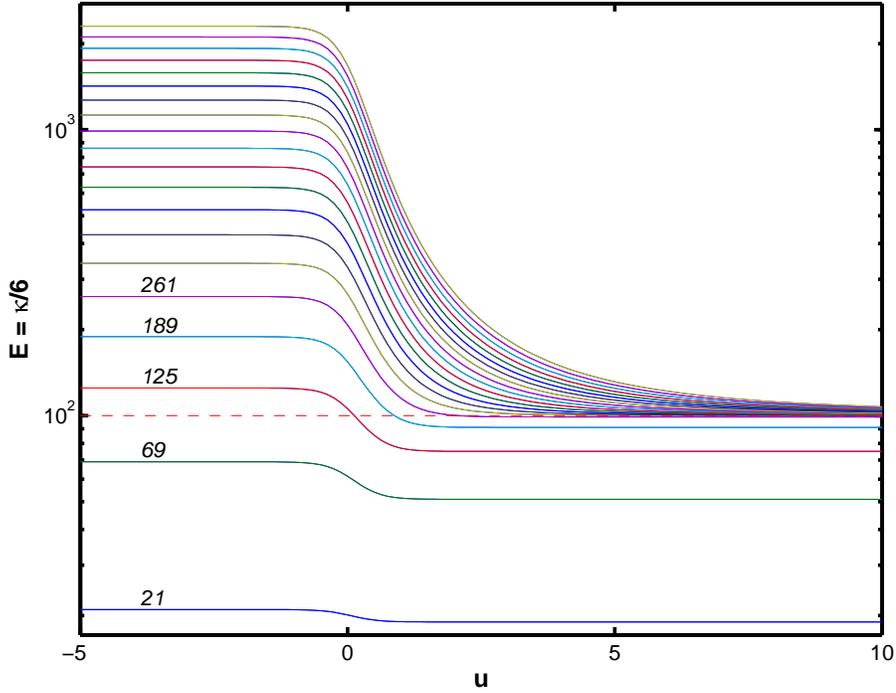,width=12.cm}}
    \caption{\sl The spectrum flow for  $m=10$; dashed line shows the continuum threshold.}
  \end{center}
\end{figure}

\begin{figure}[ht] 
\begin{center}
    \begin{picture}(160,140)(0,0)
      \DashLine(72,40)(124,40){5}
      \Line(-32,8)(4,40)
      \Line(-32,72)(4,40)
      \Line(124,40)(148,40)
      \Line(148,40)(184,40)
      \Line(184,40)(220,8)
      \Line(184,40)(220,72)
      \Line(4,40)(48,40)
      \Line(48,40)(72,40)
      \Text(-32,84)[c]{$0$}
      \Text(-34,-2)[l]{$1$}
      \Text(148,54)[c]{\mbox{N-3}}
      \Text(184,54)[c]{\mbox{N-2}}
      \Text(210,-4)[l]{\mbox{N-1}}
      \Text(210,84)[l]{\mbox{N}}
      \Text(4,54)[c]{$2$}
      \Text(48,54)[c]{$3$}
     \Text(-52,72)[r]{$MR\cosh\beta$}
      \GCirc(-32,72){4}{.5}
      \BCirc(-32,8){3}
      \BCirc(148,40){3}
      \BCirc(184,40){3}
      \BCirc(220,8){3}
      \BCirc(220,72){3}
      \BCirc(4,40){3}
      \BCirc(48,40){3}
      \end{picture}\label{fig:1}
  \end{center}
\caption{\hfil The extended Dynkin diagram for the {\sl Sausage} model.}
\end{figure}
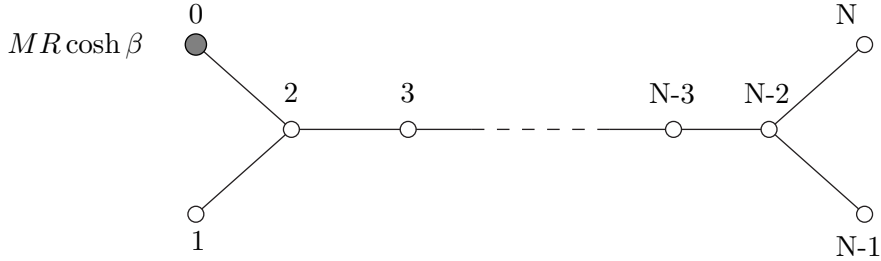

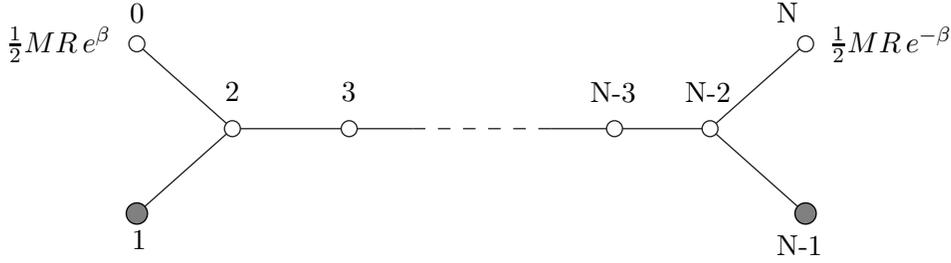
\begin{figure}[ht] 
\begin{center}
    \begin{picture}(160,100)(0,0)
      \DashLine(72,40)(124,40){5}
      \Line(-32,8)(4,40)
      \Line(-32,72)(4,40)
      \Line(124,40)(148,40)
      \Line(148,40)(184,40)
      \Line(184,40)(220,8)
      \Line(184,40)(220,72)
      \Line(4,40)(48,40)
      \Line(48,40)(72,40)
      \Text(-32,84)[c]{$0$}
      \Text(-34,-2)[l]{$1$}
      \Text(-42,72)[r]{$\half MR\,e^\beta$}
      \Text(148,54)[c]{\mbox{N-3}}
      \Text(184,54)[c]{\mbox{N-2}}
      \Text(210,-4)[l]{\mbox{N-1}}
      \Text(210,84)[l]{\mbox{N}}
      \Text(230,72)[l]{$\half MR\,e^{-\beta}$}
      \Text(4,54)[c]{$2$}
      \Text(48,54)[c]{$3$}
      \BCirc(-32,72){3}
      \BCirc(148,40){3}
      \BCirc(184,40){3}
      \BCirc(220,72){3}
      \BCirc(4,40){3}
      \BCirc(48,40){3}
      \GCirc(-32,8){4}{.5}
      \GCirc(220,8){4}{.5}
    \end{picture}
\end{center}\label{fig:2}
\caption{The extended Dynkin diagram for the SM(\ref{eq:SM})}
\end{figure}
\noindent
\section{Matching to \tba\  data}\label{sec:matching-smallsf-tba}
As it was already mentioned in Introduction for the special values of
parameter
\begin{equation}
\frac{\nu }{4\pi }=\frac{1}{N};\quad N=3,4,...  \label{nun}
\end{equation}
the effective central charge of the SM~(\ref{eq:SM}) admits the exact
({\sl i.e.} to all loops) calculation using the \tba\ equations. These
equations can be derived from the factorized scattering theory of
right ($r$) and left ($l$) moving massless particles which form the
spectrum of our SM. The factorized scattering theory of massless
particles is characterized by two-particle $S$-matrices $S_{rr}(\beta
),S_{ll}(\beta ),$ and $S_{rl}(\beta )$ where $\beta $ is the relative
rapidity of scattering particles.  We suppose to discuss in more
details this scattering theory (as well as perturbed CFT approach to
our SM) in another publication. Here we only note that each of these
three $S$-matrices coincides formally with the $S$-matrix for the
massive particles in $Z_{N}$ parafermionic CFT perturbed by the
parafermionic operators. This $S$-matrix is described in details in
Refs.\cite{fateev91a, fateev91b}.  This scattering theory results in
\tba\ equations which form the common system of $\ N+1$ coupled
non-linear integral equations for $N+1$ functions $\varepsilon
_{a}(\beta )$ of rapidity variable $-\infty <\beta <\infty $.  The
\tba\ system has the form:
\begin{equation}\label{eq:tba}
  \rho_a(\beta) = \varepsilon_a + \frac{1}{2\pi}
  \int\sum_{b=0}^N
  \varphi_{ab}(\beta-\beta')\,\log(1 + e^{-\varepsilon_a(\beta')}) 
\, {\rm d}\beta'\;,
\end{equation}
where $\varphi_{ab}(\beta) = \frac1{2\pi}{\mathcal
  I_{ab}}/\cosh\beta$, ${\mathcal I}$ being the incidence matrix of
the extended affine $D_{N}$ Dynkin diagram (see Fig.~7) and the source
term 
\begin{equation}
\rho _{a}(\beta )=\half\, RM\exp (\beta )\,\delta _{a0}+\half\,RM\exp
(-\beta )\,\delta _{aN}.  \label{sour}
\end{equation}
The effective central charge can be calculated as
\begin{equation}
c(R)=\frac{3}{\pi ^{2}} \int \sum\limits_{a}\rho _{a}(\beta )\log (1+\exp
(-\varepsilon _{a}(\beta ))\,\diff\beta .  \label{eq:cchh}
\end{equation}
The incidence matrix of these \tba\ equations is similar to that for the sausage
model (see Fig.~6), the only difference coming from the source terms. For the
massive sausage model (without topological term) $\rho _{a}(\beta )=RM\cosh
(\beta )\,\delta _{a0}$.

The \UV\  behavior of $c(R)$ is determined only by the structure of the
incidence matrix and we can find that in both cases effective central charge
approaches to the limiting value $c_{UV}=c(0)=2$ logarithmically (see
Ref.\cite{fateev93}) in agreement with Eqs.(\ref{eq:asympt},\ref{uvss}). The
analysis of \tba\  equations with source $\rho _{a}$ given by Eq.(\ref{sour})
shows that in the \IR\  limit $c_{IR}=c(\infty )=2-6/(N+2)$ coincides with
central charge of parafermionic CFT and the \IR\  corrections to this value
have a structure:
\begin{eqnarray}
c(R,\,N) &=&2-\frac{6}{N+2}+b_{2}(N)\left( \frac{N+2}{MR}\right) ^{8/(N+2)}+ 
\notag \\
&&b_{3}(N)\left( \frac{N+2}{MR}\right) ^{12/(N+2)}+...  \label{eq:cche}
\end{eqnarray}
The \IR\ asymptotics of our theory can be described by the methods of
perturbed CFT. This field theory is characterized by integrable
perturbative operator which belongs to the space of fields of the
parafermionic CFT and has the dimension $\Delta _{pert}=1+2/(N+2).$ It
gives us the possibility to calculate analytically the first
corrections to the expansion (\ref{eq:cche}).  The exact values for
the coefficients $b_{2}(N)$ and $b_{3}(N)$ are presented in
Appendix C.

The effective central charge $c(R,\,N)$ was computed using the \tba\  equations
numerically for several values of $N$ ($N=5,7,11,15,23,...$). At large $N$
the central charge is predicted to be given by
\begin{equation}
c(R,N)=2-\frac{\kappa _{0}(u)}{N+2}+O(1/N\log N)  \label{eq:cchas}
\end{equation}
To verify this we have to relate $u$ to the parameter $MR$ entering the \tba\ 
equations. Within one loop approximation we have a freedom which can be used
to fit the data in the best way. Asymptotically we expect 
$u\sim\log(N/MR)/N$, but finite $N$ corrections are present and may be
important. Empirically we find that a rather accurate choice is the
following
\begin{eqnarray}
  u &=& N_{{\rm eff}}^{-1}\,\log(N/MR)\\
  N_{\rm{eff}} &=& \sqrt{(N+2)(N-2\,\tanh(4\,\log(N/MR)))}
\end{eqnarray}
which is used to build the plot of Fig.~\ref{fig:scaling2}.  It is
quite evident that the data are increasingly well matched by $\kappa$
as $N$ increases (in the deep \UV\  or \IR\  the agreement is even better).

\begin{figure}[ht] 
  \begin{center}
    \mbox{\epsfig{file=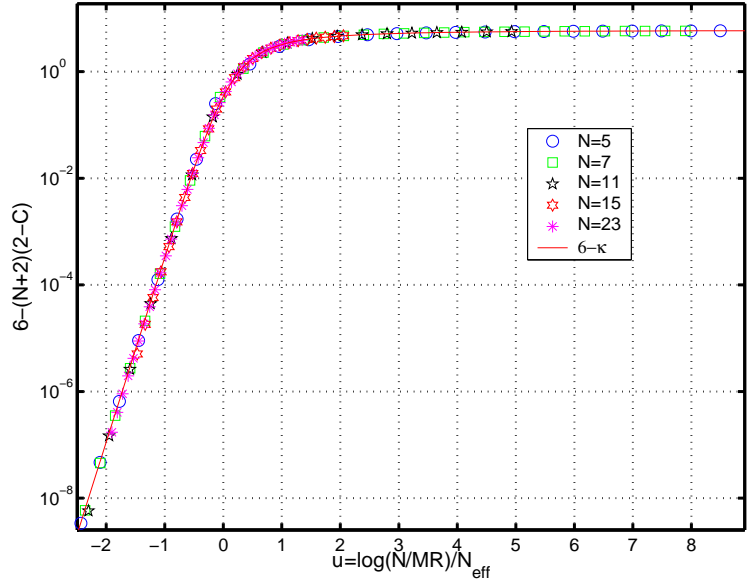,width=10.cm}}
    \caption{\sl Matching $\kappa$ to the {\sc tba} data.}
    \label{fig:scaling}
  \end{center}
\end{figure}

\begin{figure}[ht] 
  \begin{center}
    \mbox{\epsfig{file=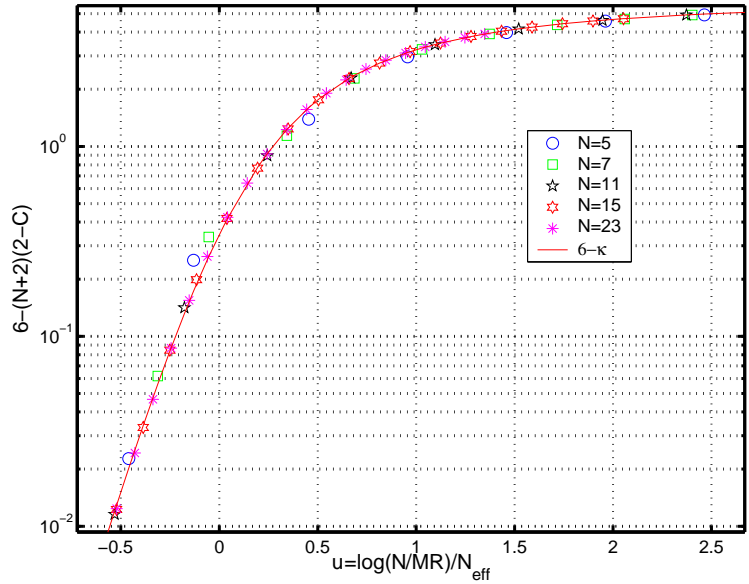,width=10.cm}}
    \caption{\sl A close--up view of Fig.9.}
    \label{fig:scaling2}
  \end{center}
\end{figure}

Finally let us note that the exact values of the coefficients $b_2(N)$
and $b_3(N)$, given in Appendix C, have been reproduced with a high
degree of accuracy by fitting the \tba\ data at various values of $N$.

\section*{Conclusions}\label{sec:conclusions}
We have found a unified treatment of the general spectral problem for
the sausage model and its variant SM~(\ref{eq:SM}). The two
formulations correspond to the same Heun equation defined on different
domains in complex plane and they are linked by a projective
transformation. The spectral function $\kappa(u)$ can be studied both
numerically by diagonalizing a discretized form of the differential
operator, or analytically by a power series expansion around the \IR\ 
point. The perturbative expansion can be pushed to high orders and it
turns out to be convergent in the whole physical domain.  Its
asymptotic behavior at high order is compatible with the leading \UV\ 
behavior, which has been computed analytically. In the ultraviolet
regime, the spectrum, besides the expected continuous component,
contains a set of bound states with angular momentum higher than one.
The scaling functions for the ground state has been compared to the
central charge computed via \tba\ equations: a very good agreement was
found, adding good evidence for the interpretation of the
SM~(\ref{eq:SM}) (at $\nu=4\pi/N$) as the field theory
describing the RG flow to the $Z_N$ parafermionic CFT.
\section*{Acknowledgments}
E.O. would like to thank R.\ K.\ Ellis and his colleagues of the Theory Group at
Fermilab for the kind hospitality he enjoyed while this work was done. We warmly
thank R.\ De Pietri who developed the symbolic code. This work was supported in
part by the EU under contract HPRN-CT-2002-00325 and grant INTAS-OPEN-00-00055.
\newpage
\section*{Appendix A}
We recall some standard results from Raileigh--Schroedinger
perturbation theory for non-degenerate levels, which should be found
in any textbook on Quantum Mechanics. Stationary perturbation theory
is most efficiently formulated as an iterative algorithm. Let $H = H_0
+ \varepsilon\, V $, $E_0$ any one of the unperturbed eigenvalues,
with eigenstate $\ket{E_0}$. Then the perturbed eigenvalue
$E(\varepsilon)$ can be expanded as
\begin{equation}
  E(\varepsilon) = E_0 + \sum_{n\ge 1} \delta_n \varepsilon^n
\end{equation}
while the eigenstate is given by a vector series in terms of auxiliary
vectors $\ket{\eta_n}$ as
\begin{equation}
\ket{E(\varepsilon)}=\ket{E_0}+\sum_{n\ge1}\varepsilon^n\ket{\eta_n}\;.  
\end{equation}
The expansion coefficients can be computed through the following
recursive algorithm:

\bigskip
{\sl Let} $\ket{\eta_0}=\ket{E_0}$ {\sl and}
$  \R_0 = \dfrac{1 - \ket{E_0}\bra{E_0}}{H_0-E_0}\,$;
{\sl then for any $n\ge1$ we have}
\begin{equation}\label{eq:algo}
  \begin{cases}
    \delta_n = \bra{E_0}V\ket{\eta_{n-1}}\;, & \\
    \ket{\eta_n} = \R_0 \left\{-V\ket{\eta_{n-1}} + \sum_{k=1}^n
    \delta_k\ket{\eta_{n-k}}\right\}\;, & 
  \end{cases}
\end{equation}

The algorithm can be easily translated in any symbolic manipulation
language. In principle we should take care of defining the matrix $V$
in an orthonormal basis; this would introduce some square roots in our
matrix Eq.~(\ref{eq:matrix}), while at the end the coefficients turn
out to be rational. Actually the following lemma tells us that we may
comfortably work with the unnormalized basis. 

\noindent
{\sf{Lemma}:} Let $H, H_0, V, \ket{E_0}, E(\varepsilon), \R, \delta_n,
 \ket{\eta_n}$ be defined as above. Let us assume further, to avoid
 any convergence problem, that $V$ be a finite-band matrix along the
 main diagonal. Let $S$ be any nonsingular Hermitian operator commuting with
 $\h_0$. Then we may substitute $V$ with $S\,V\,S^{-1}$ in the
 iterative algorithm (\ref{eq:algo}), leaving everything else
 unchanged.

\noindent
{\sf{Proof}:} Let $\delta'_n, \ket{\eta_n'}$ denote the sequence
constructed by inserting $SVS^{-1}$ into eq.~(\ref{eq:algo}). 
Since $S$ commutes with $H_0$, we must have $S\ket{E_0}=s_0\ket{E_0}$
for some non--zero $s_0$. Hence 
\begin{equation}
  \delta_n' = \bra{E_0}V\, s_0\,S^{-1}\ket{\eta_{n-1}'}\;.
\end{equation}
Let us define $\ket{\eta_n''}=s_0 S^{-1}\ket{\eta_{n-1}'}$. It's easy to check
that the sequence $\{\delta'_n, \ket{\eta_n''}\}$ satisfies the same
recursion as $\{\delta_n, \ket{\eta_n}\}$, and moreover
$\ket{\eta_0''} = s_0S^{-1}\ket{E_0}=\ket{E_0}$, hence the two sequences
must coincide.
\par\noindent
{\sf{Note}:} the statement in the lemma is strictly perturbative. $H$
and $S\,H\,S^{-1}$ could be inequivalent as operators, since $S$
and/or $S^{-1}$ may be unbounded, still they share the same
perturbative expansion.

\section*{Appendix B}
We report the explicit expression of the series expansion of $\kappa_{m,\,n}$
for small values of $(n,m)$. Of course this is just a sample; the code can
generate them to any order, the only limitation being 
computer's physical memory and time.
\begin{eqnarray*}
\kappa_{\,0,\,0} &=& 6 - \lambda^2 - \half \lambda^3 
- \tfrac{229}{720}\,\lambda^4  - 
  \tfrac{109}{480} \,\lambda^5 - \tfrac{62999}{362880} \lambda^6 - 
  \tfrac{20159}{145152}\,\lambda^7 - 
  \tfrac{299803787}{2612736000} \, \lambda^8 \\&& -
  \tfrac{72503387}{746496000} \,\lambda^9 - 
  \tfrac{173336436487}{2069286912000} \,\lambda^{10} + 
  O(\lambda^{11})\\
\kappa_{\,1,\,0} &=& 18 - 4\,\lambda  - \tfrac{16}{9}{\lambda }^2 - 
  \tfrac{352}{405}\,{\lambda }^3 - \tfrac{1972}{3645}\,{\lambda }^4 - 
  \tfrac{17408}{45927} \,{\lambda }^5 
- \tfrac{701314}{2460375}\,{\lambda }^6\\&&  - 
  \tfrac{34835788}{155003625}\,{\lambda }^7 - 
  \tfrac{204567413}{1116026100} \,{\lambda }^8
- \tfrac{1588447666493}{10358117240625}\,{\lambda }^9  \\&& - 
  \tfrac{4782354354298021}{36543437624925000} \,{\lambda }^{10} + 
  O(\lambda^{11})\\
\kappa_{\,0,\,1} &=& 54 - 24\,\lambda  - \tfrac{27}{5}\,{\lambda }^2 - 
  \tfrac{27}{10}\,{\lambda }^3 - \tfrac{23949}{14000}\,{\lambda }^4   - 
  \tfrac{34047}{28000}\,{\lambda }^5 - \tfrac{370287}{400000}\,{\lambda }^6\\&&
 - 
  \tfrac{826209}{1120000} \,{\lambda }^7 - 
  \tfrac{146655243891}{241472000000}\,{\lambda }^8  - 
  \tfrac{35351959491}{68992000000}\,{\lambda }^9 \\&& - 
  \tfrac{197594782006203}{448448000000000}\,{\lambda }^{10} + 
  O(\lambda^{11})\\
\kappa_{\,1,\,1} &=& 90 - \tfrac{196}{5}\,\lambda  -
 \tfrac{9664}{1125} \,{\lambda }^2 - 
  \tfrac{7627904}{1771875} \,{\lambda }^3 - 
  \tfrac{217386688}{79734375} \,{\lambda }^4 - 
  \tfrac{173655964928}{89701171875} \,{\lambda }^5 \\&& - 
  \tfrac{485256409132928}{329651806640625} \,{\lambda }^6  - 
  \tfrac{955858372577612032}{815888221435546875} \,{\lambda }^7 - 
  \tfrac{176847414696606187696}{183574849822998046875} \,{\lambda }^8 \\&& - 
  \tfrac{12815580494902423265411456}
   {15786519210528717041015625} \,\lambda ^9 +
  O(\lambda^{10})\\
\end{eqnarray*}

\section*{Appendix C}
In this appendix we give the exact values for the first \IR\  correction
to the levels $e_{mj}(R)$ and two first corrections to the effective
central charge or to the ground state energy $e_{0}(R)$. We express
these corrections in terms of parameter $M$ entering the \tba\ 
equations. These corrections can be calculated analytically using the
methods of integrable perturbed CFT. We consider the case $\nu /4\pi
=1/N$. In this case the \IR\  limit of SM~(\ref{eq:SM}) is described by
parafermionic CFT. The conformal dimensions of the primary fields in
this CFT are characterized by two quantum numbers $m$ and $j$
($j=|m/2|,|m/2|+1...\leq N/2$) and have a form:
\begin{equation}
\Delta _{mj}=\frac{j(j+1)}{N+2}-\frac{m^{2}}{4N}  \label{djm}
\end{equation}
It is convenient to introduce $D_{mj}=(N+2)\Delta _{mj}$. As it was
noticed in the Introduction, the limiting values of $(e_{mj}-e_{0})/24$
should coincide with $\Delta _{mj}$. With the first \IR\  correction
these values are:
\begin{equation}
\frac{(N+2)(e_{mj}(R)-e_{0}(R))}{6}=4D_{mj}-D_{mj}^{2}\frac{b_{1}(j,N)}{
j(j+1)}\left( \frac{N+2}{MR}\right) ^{4/(N+2)}+...  \label{crrr}
\end{equation}
where the coefficient $b_{1}(j,N)$ can be expressed through the function 
$g(x)=\Gamma (1+x)/\Gamma (1-x)$ and has a form
\begin{equation}
b_{1}(j,N)=\frac{2N^{2}}{(N+2)^{2}}\frac{g(\frac{1}{N+2})^{2}\,g(\frac{2j+2}
{N+2})}{g(\frac{2}{N+2})\,g(\frac{2j}{N+2})}(8\pi )^{4/(N+2)}.  \label{b1}
\end{equation}
We note that in the large $N$ limit the left hand side of Eq.(\ref{crrr})
tends to $(\kappa _{mj}-\kappa _{0})/6$. For $4D_{mj}$ and $b_{1}(j,N)$ we
have
\begin{equation}
4D_{mj}=(4j(j+1)-m^{2})(1+O(1/N)),\quad b_{1}(j,N)=2+O(1/N)\;;  \label{db1}
\end{equation}
if we now define the parameter $u$ by the relation: $\left(
  \frac{N+2}{MR}\right) ^{4/(N+2)}=$ $\exp (4u)=\lambda /(1-\lambda )$
we can see that the first term in the expansion given in
Eq.~(\ref{eq:jj}) coincides with Eq.(\ref{crrr}) at 
one loop accuracy.

For the effective central charge $c(R,N)$ (or for ground state level $%
e_{0}(R)$) it is possible to calculate analytically two further \IR\ 
corrections. Namely, the coefficients $b_{2}$ and $b_{3}$ in the
expansion (\ref{eq:cche}) are
\begin{equation}
(N+2)b_{2}(N)=\frac{N^{2}(N-2)^{2}\;g(\frac{1}{N+2})\;g(\frac{3}{N+2})\;(8\pi
)^{8/(N+2)}}{(N+4)^{2}\;(N+6)^{2}\;g(\frac{4}{N+2})\;g(\frac{-2}{N+2})^{2}}
=1+O(1/N);  \label{b2n}
\end{equation}
\begin{equation}
(N+2)b_{3}(N)=-\frac{3\,N^{4}(N-4)^{2}\;g(\frac{2}{N+2})\;g(\frac{4}{N+2})\;(8\pi
)^{12/(N+2)}}{2\,(N+4)^{4}\;(N+8)^{2}\;g(\frac{6}{N+2})\;g(\frac{-3}{N+2})^{2}}
=-\frac{3}{2}+O(1/N).  \label{b3}
\end{equation}
It is easy to see from these equations that with one loop accuracy the
\IR\ expansion for the function $\kappa _{0}(u)=6-\lambda ^{2}-\lambda
^{3}/2+...\;$ coincides with exact \IR\ expansion for the function
$(N+2)(2-c(R,N)).$ 

In the \UV\ limit the leading term for the levels
$e_{mj}(R)$ (for $j\geq m-1/2$) can also be calculated exactly and has
a form:
\begin{equation}
Ne_{mj}(R)=6m^{2}+\frac{3\,(j-m+1)^{2}\pi ^{2}N(N-2)}{2\,Z_{m}^{2}(R)}
+O(1/Z_{m}^{5})  \label{exuv}
\end{equation}
where $Z_{m}(R)=\log (8\pi (N-2)/RM)+(N-2)(\psi (1)-\psi
(\frac{m+1}{2})) + \psi(1)$. This asymptotic behavior coincides at
one loop accuracy with Eq.(\ref{eq:asympt}), where $n=j-m/2$.
\bigskip


\begin{thebibliography}{99}
\bibitem{polyakov75} A.M.\ Polyakov, Phys.\ Lett. B59 (1975) 79.
\bibitem{brezin76} E.\ Brezin and J.\ Zinn-Justin, Phys.\ Rev. B14 (1976) 3110.
\bibitem{friedan80} D.\ Friedan, Phys.\ Rev.\ Lett. 45 (1980) 691.
\bibitem{zinn-justin89} J.\ Zinn-Justin, ``Quantum field theory
  and critical phenomena'' (Oxford Science Publication, Oxford 1989).
\bibitem{fradkin82} E.\ Fradkin and A.\ Tseitlin, Ann.\ Phys, 143 (1982) 413.
\bibitem{cfmp85} C.\ Callan, D.\ Friedan, E.\ Martinec and M.\ Perry,
  Nucl.\ Phys.  B262 (1985) 593.
\bibitem{lovelace84} C.\ Lovelace, Phys.\ Lett. B135 (1984) 75.
\bibitem{candelas85} P.\ Candelas, G.\ Horowitz, A.\ Strominger and
  E.\ Witten, Nucl.\ Phys. B261 (1985) 46.
\bibitem{yang69} C.N.\ Yang and C.P.\ Yang, J. Math. Phys. 10 (1969) 1115.
\bibitem{zamolodchikov90} Al.\ B. Zamolodchikov, Nucl.\ Phys. B342 (1990) 695.
\bibitem{fateev93} V.\ A.\ Fateev, E.\ Onofri and Al.\ B.\ 
  Zamolodchikov, Nucl.\ Phys.\ B {\bf 406}, 521--565 (1993).
\bibitem{zamolodchikov86} A.\ B.\ Zamolodchikov, JETP Lett. 43 (1986) 565.
\bibitem{perelman02} G.\ Perelman, {\sl The entropy formula for Ricci flow and
geometric applications}, {\sf arXiv:math}.DG/0211159.
\bibitem{kiritis91} E.\ Kiritis, Mod.\ Phys.\ Lett., A6 (1991) 2871.
\bibitem{maldacena01} J.\ Maldacena, G.\ Moore and N.\ Seiberg,  JHEP,
2001:0107, 46 (hep-th/0108044).
\bibitem{fateev86} V.\ Fateev and A.\ B.\ Zamolodchikov, Sov.\ Phys. JETP. 63
  (1986) 913.
\bibitem{brunelli95} R.\ Brunelli and G.\ P.\ Tecchiolli, 
  J. of Comp. and Appl. Mathematics {\bf 57} (1995) 329-343.
\bibitem{heun88} K.\ Heun, Math.Ann. {\bf 33} (1888) 161-179.
\bibitem{ince56} E.\ L.\ Ince, ``Ordinary Differential Equations'',
  Dover, New York, (1956).
\bibitem{Bateman55} A. Erd\'elyi, Ed., ``Higher Trascendental
  Functions'', McGraw-Hill, New York, 1955.
\bibitem{golub96} G.\ H.\ Golub and C.\ F.\ Van Loan, ``Matrix
  computations'', 3rd Ed.n, Johns Hopkins Univ. Press (1996).
\bibitem{luke69} Y.\ L.\ Luke, ``The Special Functions and their
  approximations'', Academic Press, New York, 1969, Vol.1.
\bibitem{kahmke59} E.\ Kahmke, ``Differentialgleichungen'', Band 1, Chelsea
  Pub. Co., N.Y. 1959.
\bibitem{kato95} T.\ Kato, ``Perturbation Theory for Linear Operators'',
  Springer--Verlag, Berlin 1995.
\bibitem{reed78} M.\ Reed and B.\ Simon, ``Methods of Modern Mathematical
  Physics'', Vol.IV, Academic Press, New York, 1978.
\bibitem{Hardy} G.\ H.\ Hardy, ``Divergent Series'', Oxford U.P., 1949. 
\bibitem{dijkgraaf92} R.\ Dijkgraaf, H.\ Verlinde and E.\ Verlinde,
  Nucl.\ Phys. B 371 (1992) 269.
\bibitem{witten91} E.\ Witten, Phys. Rev. D 44 (1991) 314.
\bibitem{fateev91a} V.\ A.\ Fateev, Int.\ J.\ Mod.\ Phys.\ A16 (1991) 2109.
\bibitem{fateev91b} V.\ A.\ Fateev and Al.\ B.\ Zamolodchikov,
Phys.\ Lett.\ B271 (1991) 91.
\end{thebibliography}

\end{document}